\documentclass[final,numberedheadings]{aipproc}
\usepackage{amssymb,amsmath}
\layoutstyle{6x9} 
\begin{document}
\title[]{Physics as quantum information processing\footnote{Work presented at the conference {\em
      QCMC 2010} held on 19-23 July 2010 at the University of Queensland, Brisbane, Australia.}}
 \author{Giacomo Mauro D'Ariano}{address={{\em QUIT}
    Group, Dipartimento di Fisica ``A. Volta'', 27100 Pavia,
    Italy, {\em http://www.qubit.it}\\
    Istituto Nazionale di Fisica Nucleare, Gruppo IV, Sezione di Pavia}}
\begin{abstract} 
  The experience from Quantum Information has lead us to look at Quantum Theory (QT) and the whole
  Physics from a different angle. The information-theoretical paradigm---{\em It from
    Bit}---prophesied by John Archibald Wheeler is relentlessly advancing. Recently it has been
  shown that QT is derivable from pure informational principles. The possibility that there is only
  QT at the foundations of Physics has been then considered, with space-time, Relativity,
  quantization rules and Quantum Field Theory (QFT) emerging from a quantum-information processing.
  The resulting theory is a discrete version of QFT with automatic relativistic invariance, and
  without fields, Hamiltonian, and quantization rules. In this paper I review some recent advances on
  these lines. In particular: i) How space-time and relativistic covariance emerge from the quantum
  computation; ii) The derivation of the Dirac equation as free information flow, without imposing
  Lorentz covariance; iii) the information-theoretical meaning of inertial mass and Planck constant;
  iv) An observable consequence of the theory: a mass-dependent refraction index of vacuum. I will
  then conclude with two possible routes to Quantum Gravity.
\end{abstract}
\maketitle
\section{Introduction}
It is interesting to explore the possibility that pure information may underlie all of Physics. From
what we know, such information should be made of quantum bits, instead of classical
bits.\footnote{In Ref. \cite{support} G. Chiribella, P. Perinotti, and myself have shown how QT can
  be entirely derived from purely informational principles, i.~e. QT is indeed a kind of information
  theory. Ref. \cite{support} contains all proofs in full details: it is partly based on Ref.
  \cite{CDP2010}, and closes the axiomatization program of Ref. \cite{myCUP2009}.}  The fundamental
problem is now to establish if there is something more than qubits at the foundations of Physics,
namely if there is something more that QT in a quantum field. Can we say that a quantum field is
just a collection of (infinitely many) quantum systems, each at every ``space point'' (a Planck
cell?) unitarily interacting with a bunch of other systems?  This would mean that the quantum field
(i.~e. virtually everything) is like a giant quantum computer.  Related questions are: Does the
continuum play a fundamental role (or it is only a mathematical idealization)? Are space and time
emergent? Can we recover all physical notions (as energy, charge, inertia, relativistic covariance,
and gravitation) as features of a quantum information processing?  A positive answer would be the
realization of the dream of John Archibald Wheeler: a physical world made of informational units. In
other words: "Physics is information"---the opposite of Landaurer's dictum: "Information is
physical".  This contrast of paradigms resembles a "which came first, the chicken or the egg?''
dilemma.  But, between the two hypothesis, the Occam's Razor will definitely choose the one which is
the most theoretically parsimonious. And the Wheeler's hypothesis seems to be by far the most
economical, since it is equivalent to deriving the whole Physics from solely QT, keeping only
general methodological principles (e.~g. the assumption of locality of interactions).  And, as we
will see here, this has also the great advantage of deriving Special Relativity from QT, opening the
route to the reconciliation of General Relativity with QT.

Substituting a world made of particles in a Minkowski space-time with an evanescent cosmos made of
pure information is definitely a huge change in ontology: many physicists---especially the believers
in hidden variables---will be reluctant to adopt the new view. But the notions of particle and
space-time are themselves quite inconsistent as ontologies within the current theoretical framework,
being the particle a ``quantum state'' of the field---a subjective entity in the Bayesian
approach---and being space-time without events a ``non-being'' which, however, possesses a
property: the metric.

\section{Space-Time and Special Relativity emerge from the information processing}

The first step of the informational program is understanding how Relativity supervenes upon
information processing. In the informational approach the ``real entities''---the observational
primitives---are the {\em events}.\footnote{I like to think to events as the ``facts'' of Ludwig
  Wittgenstein's {\em Tractatus}.} We formulate a {\em theory} of the events by building up a set of
causal connections between them. Events are the primordial notion: they do not happen within
space-time, but, viceversa, space-time is a construct emerging from the network of events. The goal
is now to derive space-time endowed with relativistic covariance from the network of events. This
is in the spirit of the research line launched by Rafael Sorkin more than twenty years ago
\cite{sorkin}.

In February 2010 \cite{mypirsa,Dirac_vaxjo2009} at the Perimeter Institute I gave a ``visual proof''
of how Special Relativity emerges from the computation, showing how the Lorentz time-dilation and
space-contraction can be derived by just event-counting. In a quantum computer the events are the
unitary transformations of the gates, and the causal links are the wires connecting gates: the wires
are the ``quantum systems'', or, in other words, the ``registers'' where information is written.
Now, in a quantum computer the information can flow at the maximal speed of one gate per step
(since, otherwise it should run from the output to the input).  Moreover, due to discreteness of the
circuit, information can flow in a fixed direction only at the maximum speed: the only way to slow
down the flow is by repeatedly changing direction, as in the Dirac's Zitterbenwegung.  Synchronicity
of causally independent events/systems is up to the observer. A set of synchronous events/systems
crossing the whole circuit is a leaf of a {\em foliation}, as in the Tomonaga-Schwinger quantum
relativistic approach.  By stretching the quantum circuit, we can geometrically dispose the leaves
on parallel horizontal lines, so that the vertical axis represents the chosen synchronization time.
The stretching leaves the topology invariant, namely the quantum circuit represents the same quantum
information processing (longer/shorter wires represents the same causal connections). The Lorentz
time dilation is then obtained by considering a tic-tac of an Einstein clock made with light
bouncing between two mirrors (here corresponding to information bouncing at one gate per step
between two locations) and then counting the events occurring during the tic-tac in the two
different reference systems---the un-stretched and the stretched circuits.  Similarly for Lorentz
space-contraction. In this way space-time with its Minkowski metric is emerging by event-counting
from pure topology. To be more precise, the causal network of the quantum computation is a {\em
  dressed} topological network, namely the links have a label corresponding to different types of
causal relations, e.~g.  registers/systems of variable nature as qubits, qutrits, etc.

With Alessandro Tosini we gave an analytical derivation of the Lorentz transformations from a
dynamically homogeneous causal network \cite{Lorentz}, i.~e. a periodic network made of identical
tiles connected each other in the same way. Dynamical uniformity plays the role of the Galileo
principle, i.~e. the invariance of the physical law with the reference system---the physical law
being the causal connection-rule of the network given by a repeated tile of the pattern. In the
usual space-time description the Galileo principle is also synonymous of uniformity of space and
time: however, now the causal network captures the conventionality of homogeneity of space and
time,\footnote{The problem of conventionality of synchronization was first raised by Reichenbach. He
  sayed that the speed of light can be only measured on a closed path, since there is a circular
  argument in the measurement of the one-way speed of light. Indeed, to synchronize distant clocks
  we need to know a speed, and to measure a speed we need synchronized clocks. As a matter of fact,
  all measurements of light speed either concern the two-way averaged speed---see Fizau's or the
  Michelson-Morley's---or assume synchronization of clocks in different positions---as in the
  Roemer's measurement. Notice that in the end, also measurements of distances need synchronized
  clocks.} which has been an old debated issue in Special Relativity \cite{Brown,JammerSync}.
Thus a dynamically homogeneous causal network exactly represents the Galileo principle stripped of
conventions. For the explicit derivation of the Lorentz transformations from the causal network the
reader is addressed to Ref.  \cite{Lorentz}. here I just want to emphasize that the whole procedure
is made only via event-counting. Also, an essential ingredient is a coarse-graining of events,
corresponding to the usual rescaling of Minkowski space-time, made in order to restore the symmetry
between the observers.

\section{Dirac equation as information flow and the meaning of inertial mass and Planck constant}  
The informational paradigm exhibits its full power when deriving the Dirac equation without Special
Relativity.\footnote{This part of the proceedings has been already published in more details in
  Refs. \cite{Dirac_PRL, Dirac_vaxjo2010}, which were written few months after QCMC, and which,
  however, better represent the spirit of the original talk.} Indeed, we will see now how the free quantum field is
just the description of the free propagation of quantum information along the circuit. We stick to a
single "space" dimension, but everything can be generalized to more dimensions.

On the quantum circuit information can flow only in two directions, and if it does not change
direction, it must flow at the maximum speed of one gate per step.  Everything in the circuit is
digital: the metric of space-time is adimensional, since it is made with event-counting. If we want
recover our usual notion of space-time, we need to introduce conversion units. These can be
interpreted as minimal space-distance $a$ (called {\em chorus}) and a minimal time-distance $\tau$
(called {\em chronon}) between events. The maximal speed of the information flow will be then
$c=a/\tau$, and if we take the ratio $a/\tau$ as a universal quantity we must choose $c$ equal to
the speed of light. We describe the information flows in the two directions, right and left, by
two fields $\phi^+$ and $\phi^-$.  In equations one has
\begin{equation}\label{nozigzag}
\widehat\partial_t
\begin{bmatrix}\phi^+\\\phi^-\end{bmatrix}=
\begin{bmatrix}c\widehat\partial_x & 0\\ 0&
  -c\widehat\partial_x\end{bmatrix}\begin{bmatrix}\phi^+\\\phi^-\end{bmatrix}, 
\end{equation}
where the caret on the derivative denotes that it is a finite-difference and can generally involve
more than a single step (it turns out that we need at least two forward and two backward steps).
Now, as already mentioned, the only way of slowing-down the information flow is to have it changing
direction repeatedly. A constant average speed corresponds to a simply periodic change of direction,
which is described mathematically by a coupling between $\phi^+$ and $\phi^-$ with an imaginary
constant.  Upon denoting by $\omega$ the angular frequency of such periodic change of direction, we
have
\begin{equation}\label{zigzag}
\widehat\partial_t
\begin{bmatrix}\phi^+\\\phi^-\end{bmatrix}=
\begin{bmatrix} c\widehat\partial_x & -i\omega\\ -i\omega &
  - c\widehat\partial_x\end{bmatrix}\begin{bmatrix}\phi^+\\\phi^-\end{bmatrix}. 
\end{equation}
The slowing down of information propagation can be considered as the {\em informational meaning of
  inertial mass}, and $\omega$ represents its value. Eq.  (\ref{zigzag}) is nothing but the Dirac
equation (without spin): this means that the quantum-information processing corresponding to pure
information transfer simulates a Dirac field---the periodic change of direction being the
Zitterbewegung. Notice how Eq.  (\ref{zigzag}) has been derived only as a general description of a
uniform information transfer, without requiring Lorentz covariance.

The analogy with the Dirac equation leads us to write the coupling constant in terms of the Compton
wavelength $\lambda=c\omega^{-1}=\hbar/(mc)$. This allows us to establish the relation
$m=(c^{-2}\hbar)\omega$ between $m$ and $\omega$, providing an informational meaning to the
Planck constant $\hbar$ as the conversion factor between the informational notion of inertial mass
in sec${}^{-1}$ and its customary notion in Kg. Also notice how equivalence between the two notions
of mass corresponds to the Planck quantum expressed as rest energy.

In the quantum circuit describing the free information flow (Eq. (\ref{zigzag}) with renormalization
of velocity: see Sect. \ref{s:massindex}) all gates perform unitary transformations that are far
from the identity (they are close to unitary swaps). There is therefore no notion of Hamiltonian as
infinitesimal generator of the field evolution. However, remarkably the field Hamiltonian is
recovered as an effective one emerging from the quantum gates, and defined in terms of differences
of local unitary transformations \cite{Dirac_PRL}. Also, at least for one space dimensions, it can
be proved that the gates act only on local algebras of qubits (or harmonic-oscillators in the Bose
case), and the quantum field can be thus eliminated \cite{Dirac_PRL}.  It is easy to see that the
sum of $\sigma_n^z$ on all qubits ($a_n^\dag a_n$ for oscillators) is conserved, and we can chose
the vacuum with all qubits/oscillators in the $|0\rangle$ state. The quantization rules then
re-emerges from single-qubit states (corresponding to the so-called pulse-position-modulation (ppm)
information encoding), whereas the classical trajectories come out as {\em typical paths} of the
information flow on the circuit.

\section{Observational consequences: mass-dependent refraction index of  vacuum.\label{s:massindex}}
When introducing the coupling between the left and right information flows we didn't consider the
possibility that the coupling may affect the speed $c$ of the field in Eq. (\ref{zigzag}). This
means that, due to the coupling, the speed in the Dirac equation would be no longer equal to the
maximal speed of one gate per step. As shown in Ref. \cite{Dirac_PRL}, the renormalization of the
speed from $c$ to $\zeta c$ with $0<\zeta<1$ occurs due to unitariety of the evolution, and is
independent on the specific dynamical structure of the circuit, namely it holds for flows more
general than the uniform, e.~g.  including drift terms. For the Dirac equation one has the bound
\cite{Dirac_PRL}
$\zeta\leq\sqrt{1-\left(\frac{2a}{\lambda}\right)^2}=\sqrt{1-\left(\frac{m}{M}\right)^2}$, where
$M=\hbar/(2ac)$. Thus the information flow stops completely at field-mass $m=M$. $M$ is the Planck
mass if we choose $2a$ equal to the Planck length, namely with two qubits of information per Planck
length (in the spin-less Fermi case). This should be compared with the mini black hole of the
holographic principle (with Schwartzild radius equal to $\lambda$), with information of 1 bit per
Planck cell on the surface of the black hole. One may speculate if the information halt at $m=M$ has
any relation with the holographic principle.

\begin{figure}[ht]
\includegraphics[height=1.5in]{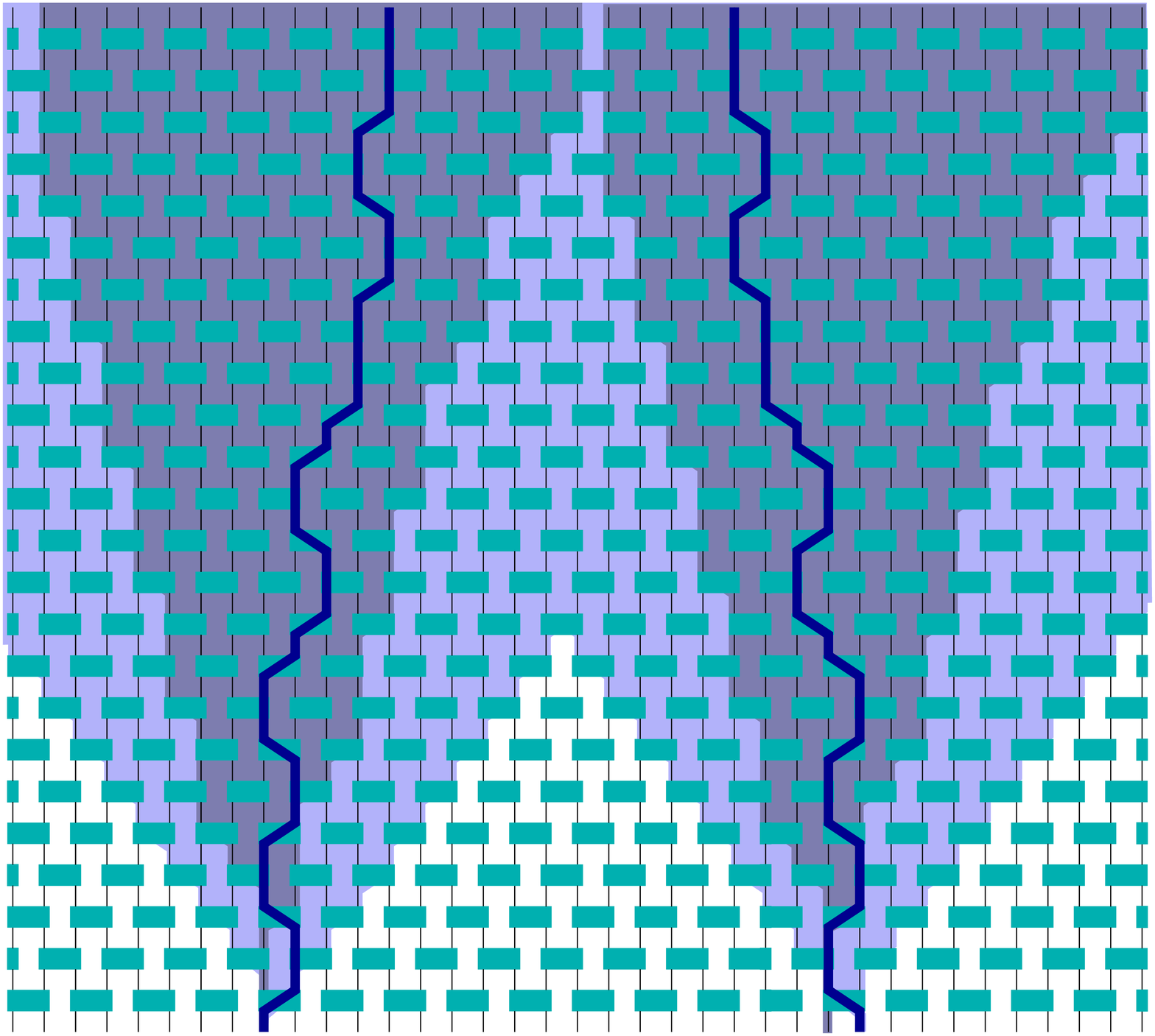}\;\;
\includegraphics[height=1.5in]{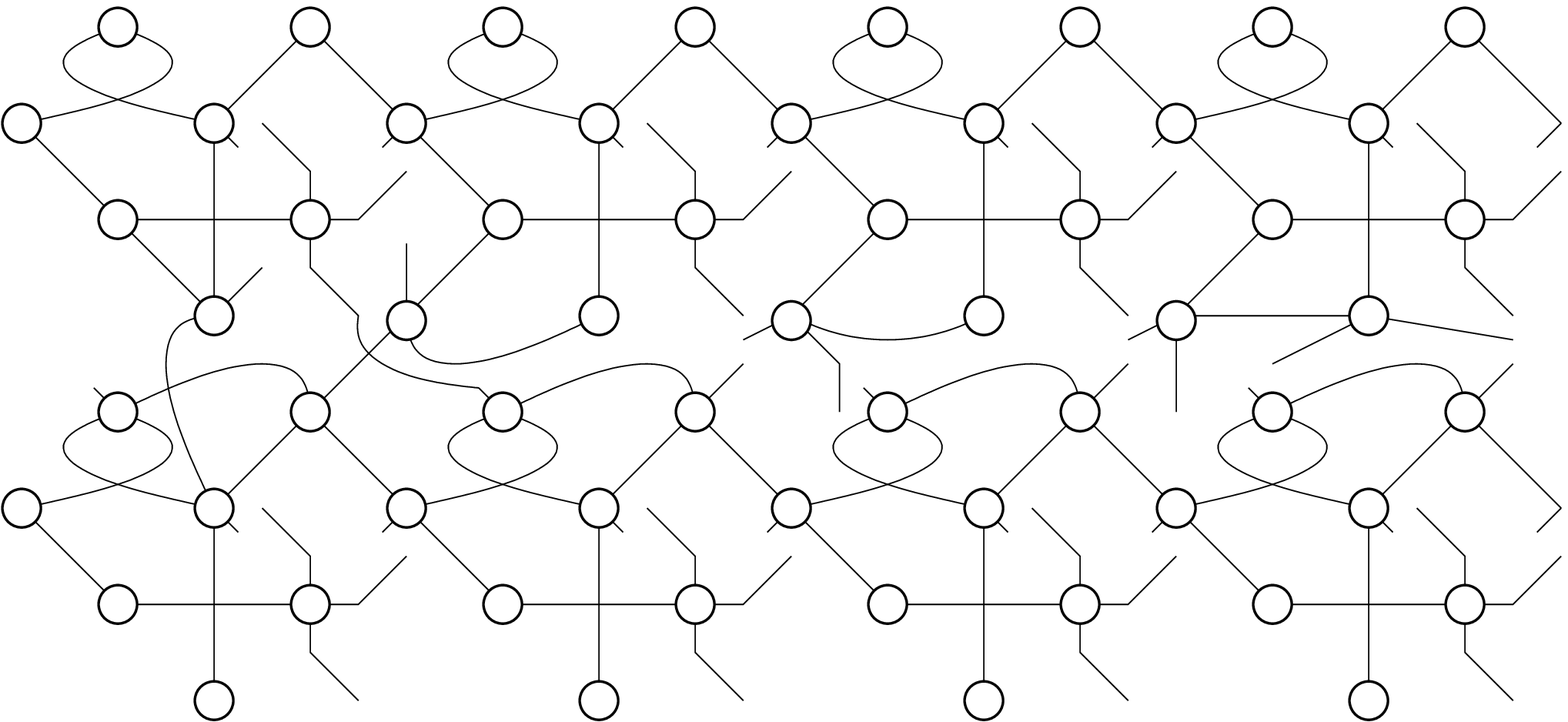}
\caption{\rm Two alternative hypothesis in an informational approach to Quantum Gravity.  {\bf
    Left:} Gravity is a quantum effect detectable at the level of pure quantum information flow (the
  figure is not a simulation, but serves only as an illustration: see text). {\bf Right:} a
  quantum computational network is considered with dynamical causal relations, in a
  third-quantization approach (see text).\label{fig:2gravity}}
\end{figure}

The last considerations naturally raise the question: Where is gravity? What is the informational
meaning of gravitational mass? At the present stage of the informational program I can just
speculate about possible future lines of research. The first possibility is that we believe in a
strong version of the equivalence principle, i.~e. that inertial and gravitational masses are
actually the same informational entity. Then, gravity must be a quantum effect, similarly to the
case of the {\em induced gravity} of Andrei Sakharov \cite{sakarov,visser}. This means that we may
suppose that gravity should manifest at the level of free field, or, in other words, we should see
gravitational effects already in the quantum information processing of pure information flow. Why
then this effect does not occur in the usual Dirac theory? In the quantum circuit there may be an
effect, although non visible in the usual free QFT, because the behavior of the solutions of the discrete
Dirac equation differ from the continuous case, especially due to the possibility of perfect
localization of information. For example, one may consider the evolution of a state with two qubits
up (over a vacuum of qubits down). Due to Zitterbewegung the qubit will spread in time in a
superposition cone as in Fig.  \ref{fig:2gravity}. The gravitational interaction may result e.~g.
from the cones of the two qubits meeting at some gate, where they interact, producing a distortion
in the average paths of information.  A second possible line of research (if we do not believe in
the strong equivalence principle) is to consider a quantum computational network with dynamical
causal connections. The causal connections may be ``programmed'' by another parallel circuit (the
circuit version of a gauge-field), or we may even want to relax causality of QT and consider a sort
of third-quantization in which the causal links---i.e. the {\em systems}---become themselves quantum
states of some higher-level systems.

\subsubsection*{Acknowledgments} 
I thank Seth Lloyd, Rafael Sorkin, and Alessandro Tosini for very interesting discussions and
suggestions. I'm grateful to researchers of the Frascati Labs, Chicago Fermilab, and Perimeter
Institute for useful feedback, encouragement, and hospitality.

\end{document}